# Influence of Solution Parameters on Phase Formation and Morphology of Electrospun Poly(vinylidene fluoride) Nanofiber

*Anshika Bagla [1], Chhavi Mitharwal [1], François Rault [2], Fabien Salaün [2], Supratim Mitra [1\*]*

[1]*Department of Physical Sciences, Banasthali Vidyapith, Rajasthan 304022, India*

[2] *Univ. Lille, ENSAIT, ULR 2461 - GEMTEX - Génie et Matériaux Textiles, F-59000 Lille, France*

*\*Correspondence: supratimmitra@banasthali.in*

## Abstract

Poly(vinylidene fluoride), PVDF nanofibers were prepared using electrospinning method and the influence of electrospinning parameters such as PVDF concentration, DMF:Acetone ratio on the formation of different phases (α, β, and γ) and desired morphology were investigated. A detail analysis of X-ray diffraction (XRD) and Fourier-transform infrared (FTIR) spectrometer data by deconvoluting the peaks were carried out for assigning and accurately identifying the peaks associated to a particular PVDF phase. SEM micrographs further helped in understanding the mechanism of desired β-phase as well as in obtaining a condition for bead-free nanofibers. Interestingly, the mechanism of formation of phases are found mostly governed by the balance between surface tension and viscosity which is controlled by PVDF concentration and acetone content in the solution. A PVDF concentration of 20% (w/v) and DMF:Acetone ratio 1:1 was found suitable for maximum β-phase in a completely bead-free nanofibers. The methodology and understanding of underlying mechanism of desired phase and morphology suggest a guideline for optimization of β-phase and bead-free fibers in PVDF-based nanocomposite as well.

**Introduction**

Polyvinylidene fluoride (PVDF) is the most considerable electroactive (EA) fluoropolymer that have been studied in last couple of decades by reason of their attractive piezoelectric and ferroelectric properties, which are utilized in a variety of applications including sensors, actuators, and nanogenerators [1-6]. PVDF nanofibers made by electrospinning have recently gained much interest to researchers as a multifunctional fiber production method that has been applied in a range of commercial industries, including membrane and filter technologies. [7-11], tissue engineering and drug delivery in healthcare, fuel cell, battery, supercapacitors and nanogenerators in energy, as well as defense and security [12-14]. Many of these applications demand presence of maximum EA-phase and the control over the fiber morphologies [12, 15, 16]. Therefore, the challenge has always been to control the electrospinning parameters to obtain a maximum EA-phase and desired morphologies at the same time. Furthermore, PVDF is a semi crystalline polymer known to exist in EA polar β-phase in all-*trans* (*TTTT*) and less polar γ-phase in tri-*trans gauche* (*TTTG$^+$TTTG$^-$*), and non-polar α-phase in alternate *trans gauche* (*TG$^+$TG$^-$*) conformations [17, 18]. Further, the amount of EA-phase in PVDF is largely determined by the β-phase content, microstructure and degree of crystallinity, which are highly influenced by the processing conditions. Thermal, mechanical, and poling treatments, as well as solution casting method utilizing various solvents, are commonly used to obtain the most desirable β-phase from an existing α-phase [19]. However, these procedures are laborious and time consuming, even not always very effective as they can produce several coexisting crystalline phases [15, 20-22]. On the other hand, electrospinning provides *in-situ* mechanical stretching and electrical poling to obtain β-phase from polymeric solution [20, 23, 24]. Despite the fact that electrospinning is a convenient and versatile method, the actual dynamics and mechanism consisting of several process steps starting from electrification of polymeric solution to jet initiation followed by elongation and solidification of nanofibers, are not well understood [25]. Moreover, several studies have been focused on effect electrospinning parameters such as solution (viscosity [10, 26-28], conductivity, surface tension [10, 27-31], molecular weight [26, 31-34], concentration [10, 23, 34-36]), process (applied field [10, 33, 34, 37, 38], tip-collector distance [10, 31, 34, 39-41], flow rate [10, 28, 31, 38, 42], needle diameter), and ambient parameters (temperature [29, 31, 43], humidity [44-46]) to understand these process steps and their influence on attainment of a desired EA-phase and morphology of PVDF nanofibers. However, there exists some inconsistencies in these results, which are due to the obvious reason that it contains a significant number of interdependent electrospinning parameters and a complex interaction between these key parameters determines the phase and morphology of the resulting nanofibers [47, 48]. Further, inaccurate identification of XRD and FTIR peaks for different coexisting phases (e.g., α, γ in XRD; β, γ in FTIR) also results in inconsistencies in phase quantification [15, 49]. Most importantly, comprehensive data and understanding of how the various electrospinning parameters influence the crystallinity, β-phase content and morphology, as well as a correlation among them, are still lacking.

In an earlier work, authors have summarized the influence of various electrospinning parameters such as solution (molecular weight, concentration, solvent type), process (applied voltage, flow rate, needle size, tip-to-collector distance), and ambient (relative humidity, temperature) parameters and established some of the major parameters for tailoring the nanofibers [50]. As a result, and also considering the fact that, varying each electrospinning parameters simultaneously is very much difficult, this work focused on investigating the influence of some of the important solution parameters on various stages of electrospinning nanofiber formation, as well as the morphology and phases in the resulting PVDF nanofibers. For this, we have varied the PVDF concentration and ratio of the binary solvent (DMF:Acetone) used for the preparation of PVDF nanofibers. Different phases obtained at varied electrospinning parameters were characterized using X-ray diffractometer (XRD) and

Fourier-transform infrared spectroscope (FTIR), which helped to identify those individual phases present in the PVDF nanofibers and quantify them. Further, using a Scanning electron microscope (SEM), morphology of the resulting nanofibers was investigated and explanation for the mechanism of different phase formation due to varied electrospinning parameters and a correlation among them to find the most suitable condition was provided. The findings allowed us to find a suitable combination of PVDF concentration and DMF:Acetone ratio for obtaining a maximum β-phase in completely bead-free nanofibers, which could provide a guideline for nanofiber preparation in pristine as well as PVDF based nanocomposites.

**EXPERIMENTAL SECTION**

**Preparation of PVDF nanofibers**

For the preparation of electrospun nanofibers, it is considered that, ambient parameters (temperature, and humidity) do not change significantly during deposition and keeping standard temperature 25 $^o$C and relative humidity 40-45% [50]. A PVDF solution of concentration 15, 20, and 25% (w/v) were prepared by dissolving PVDF powder (Molecular weight of 1,80,000; 99.9% purity, Sigma-Aldrich, India) in a mixture of DMF/acetone (Both were purchased from Sigma Aldrich with 99% purity) taken in a ratio of 1:1, 1.5:1, and 2:1 by continuously stirring at 60 $^o$C on a magnetic stirrer. Sample ids are assigned hereafter as *PVDF concentration_DMF:Acetone* (e.g., 15_1:1; 15_1.5:1; 15_2:1; 20_1:1; 20_1.5:1; 20_2:1; 25_1:1; 25_1.5:1; 25_2:1). First, the solution was filled in a 10 ml plastic syringe with a blunt edge needle (23G) of inner diameter 0.33 mm. The syringe was placed in an infusion pump (NE-300, New Era, USA) to maintain a constant flow rate of 1 ml/h. A dc voltage of 18 kV was applied between the needle and the metallic stationary collector that was kept 12 cm apart. The nanofiber mat was directly collected on an Al (aluminium) foil and dried at 60 $^o$C in oven for 6 h.

**Characterization of PVDF nanofibers**

The phase formation of the PVDF nanofiber mat was confirmed using an X-ray diffractometer (Bruker D8 Advance, Germany) pattern and Fourier transform infrared (FTIR) (PerkinElmer, USA) spectra data. Morphology was studied using a scanning electron microscope (SEM) (Tescan MIRA 3 MUG FEG, Czech Republic). The diameter of the fibers was measured using Image J (USA) software from the SEM micrographs. For accurately identify the peak position of different phases in PVDF nanofiber, first, the peaks that obtained from XRD data for the 2θ range 10-30º were deconvoluted. The deconvoluted XRD peaks were then indexed [15, 51] and assigned phases by comparing them to the phases obtained from FTIR peaks [49, 52]..

**Results and discussion**

Once a polymeric solution begins to form nanofibers, the goal is to achieve consistent and controllable fiber diameter, defect-free fiber surfaces, and collectable continuous single fiber. As a result, to achieve the desired morphology and phase of PVDF nanofibers, various electrospinning parameters are varied. Due to the large number of parameters, however, some of them must remain constant. Furthermore, it is believed that, during the slow acceleration of pre-solid state fibers near the needle tip are attributed to electrification and jet initiation that is controlled by both the process parameters (applied voltage, tip-to-collector distance and flow rate, needle diameter) and solution parameters (viscosity, surface tension, conductivity, concentration, molecular weight), whereas during fast acceleration solid state fibers near the collector is attributed to elongation and solidification is controlled by mainly solution parameters [25, 28]. Recognizing that solution parameters have an impact throughout the electrospinning process, we kept process parameters constant while varying the most relevant solution parameters such as solvent and concentration. In this study, the solvent is a mixture of polar *N*, *N*-dimethylformamide (DMF), and acetone in various ratios. Acetone, which is highly volatile, can reduce the viscosity and surface tension of a solution while increasing the rate of evaporation. When PVDF is dissolved in a binary solvent, the viscosity of the solution is proportional to the concentration of the PVDF, whereas surface tension is completely controlled by the solvent composition and has little to do with concentration. Another important parameter that influences viscosity is molecular weight, which was taken 180,000 in this study. The particular molecular weight is chosen because very low molecular weight results in highly beaded fibre and very high molecular weight can clog the needle. The conductivity of the polymeric solution is not changed by adding any conducting salt, and it is also not expected to change significantly in our study due to very small differences in conductivities of DMF and acetone (conductivity of acetone and DMF is 0.74 and 0.92 µS/cm, respectively [53]).

**Phase identification and quantification**

Figure 1 shows the XRD patterns of all the samples with varying PVDF concentrations (15, 20, 25 %w/v) and DMF:Acetone ratios (1:1, 1.5:1, 2:1) that were collected for the 2θ range 10-90°. As the PVDF concentration is increased from 15 to 25%, a predominating α-phase characterized by peaks from (100) plane at 2θ ≈ 18.3° and (021) at 2θ ≈ 26.6° gradually disappeared while a weak broad peak at around 40.7° from (002) plane exists for all samples. A strong and sharp peak at around 2θ ≈ 20.5° progressively appeared at higher concentrations of PVDF is attributed apparently from the dual characteristics of β from (110/200) or γ-phase from (110). Hence, it must be clarified as this problem regarding identification of β and γ-phase persisted in the literature for long back [37, 40, 54, 55]. Furthermore, two relatively weak characteristic peaks of β-phase appeared at 2θ ≈ 36.2 and 56.0° from (101) and (221) plane respectively. These observations imply that as PVDF concentration increases, the formation of an EA-phase becomes more favorable compared to α-phase. Further, quantification of the

EA phase, particularly individual β and γ-phases, as well as the degree of crystallinity in distinct PVDF samples, becomes critical to understanding its piezoelectric properties. This requires a peak deconvolution to identify each phase separately and their contribution in calculating the overall crystalline phase of the PVDF. The deconvoluted peak profile of all the samples for the 2θ range 10-30º with varying concentration and DMF:Acetone ratios is depicted in Figure 2. The presence of the exclusive diffraction peak from the α(100) plane at 17.7º revealed a pre-dominating α-phase in 15_2:1 and 15_1.5:1, whereas it was absent in 15_1:1. On the other hand, α(100) still present in all 20% but rather weak compared to 15%, while it was completely absent in all 25% PVDF samples. The other exclusive diffraction peak for α-phase from α(021) at 26.6º is evident in all 15%, but gradually diminished for 20%, and disappeared in 25_1:1 and 25_1.5:1 except for extremely small peak in 25_2:1. This proved that, α-phase is more dominating phase in low PVDF concentration, however as the PVDF concentration is increased the α-phase fraction is reduced. Moreover, it was also found that, when the DMF:Acetone ratio is decreased for any particular PVDF concentration, α(100) and α(021) peak intensities are also decreased, indicating a reduction in the dominant α-phase fraction. Furthermore, the other characteristics peaks of α-phase, α(020) at 18.3º and α(110) at 19.9º are clearly visible in 15_2:1 and 15_1.5:1. However, for 15_1:1, the diffraction peak α(020) has emerged as mixed phase with γ and appeared as α,γ(020) at slightly higher 2θ value of 18.5º, while the peak α(110) is attributed to γ(110) that appeared also at little higher 2θ value of 20.3º, since both exist in monoclinic phase [24, 51]. This further confirms that as DMF:Acetone ratio decreases, the formation of an EA-phase becomes more favorable to that of a pre-dominant α-phase. Similar result has been obtained in both 20 and 25% also, where the diffraction peak α(020) has appeared as a mixed α + γ phase at 18.5º as α,γ(020). The peak α(110), on the other hand, has appeared as mixed α + γ phase as α,γ(110) at 20.04º in 20%, while, it appeared as γ(110) at 20.3º in 25% [56]. This is also indicating a significant reduction in α-phase in both 20 and 25% PVDF. This remained always a challenging task to many authors to identify γ(110) as it appeared in mixed phase as α,γ(110), even sometimes mistakenly reported as β(110/200) [15, 37, 40, 55]. Furthermore, the only characteristics diffraction peak for β-phase, β(110/200) has appeared at 20.7º for all samples (15, 20, 25%), however becomes more sharper and stronger as PVDF concentration is increased irrespective of DMF:Acetone ratio.

Thus, it can be concluded that when the DMF:Acetone ratio is low (1:1), the formation of an electroactive γ and β-phases to that of a predominant α-phase is more favorable for a fixed PVDF concentration, while for a fixed DMF:Acetone ratio, as the PVDF concentration increases, more electroactive γ and β-phases are developed. As a result, a quantification of each phase would be critical at this stage to investigate the influence of concentration and DMF:Acetone ratio on the phases as sometimes different phases (α, γ) exhibit similar peaks in PVDF nanofibers [15]. A common way several authors have quantified relative proportion of β-phase from XRD data is sum of the intensity ratio of peaks associated to β-phase to those α, β, and γ-phases and is given by [57-59],

$$\frac{I_\beta}{I_\alpha+I_\beta+I_\gamma} = \frac{I_{(200/110)}+I_{(101)}+I_{(221)}}{I_{(100)}+I_{(020)}+I_{(110)}+I_{(021)}+I_{(002)}+I_{(200/110)}+I_{(101)}+I_{(221)}} \quad \text{Equation (1)}$$

where, $I_\alpha$, $I_\beta$, $I_\gamma$ are peak intensities for α, β, and γ-phases respectively and planes (100), (110), (021), (002) associated to α-phase, (200/110), (101), (221) planes to β, (020), and (110) planes to (α + γ) and γ-phase respectively. The relative proportion of β-phase are summarized in Table 1. It can be seen that the increase in the PVDF concentration increases the formation of overall β-phase (in the range 30.0-38.2% for 15%; 40.9-59.2% for 20% and 53.7-51.6% for 25%) in a particular PVDF concentration. Moreover, in 20% PVDF and DMF:Acetone ratio 1:1 (20_1:1) a maximum β-phase 59.2% is obtained. It is also worth mentioning here that, as the quantification of each crystalline phases in PVDF depends on the overall crystallinity of the sample, the degree of crystallinity ($\chi_c$) was calculated from,

$$\chi_c = \frac{\sum A_{cr}}{\sum A_{cr} + \sum A_{amr}} \times 100\% \qquad \text{Equation (2)}$$

where, $\sum A_{cr}$ and $\sum A_{amr}$ are the sum of integral area of crystalline peaks and amorphous halo respectively [60]. The overall crystallinity, $\chi_c$ (%) is found to increase with the concentration, however no systematic variation found due to addition of acetone in the solvent for a particular concentration of PVDF as can be seen in Table 1.

FTIR spectra of all the samples with varying PVDF concentrations (15, 20, 25 %w/v) and DMF:Acetone ratios (1:1, 1.5:1, 2:1) were collected for the frequency range 450-1500 cm$^{-1}$. The peaks corresponding to vibrational bands associated to each PVDF crystalline phases (α, β, γ) are labeled as shown in Figure 3. However, some of the common peaks in the range, 876-885, 1067-1075, 1171-1182, 1398-1404 cm$^{-1}$ are not considered as they appear in mixed phases. The predominant α-phase is characterized by two exclusive absorption peaks at 614, 763 cm$^{-1}$ and other characteristics peaks at 489, 532, 796, 976 cm$^{-1}$. These peaks are comparatively more intense in 15%, and gradually faded as PVDF concentration is increased to 20 and further to 25% indicating formation of EA-phases (β, γ) compared to predominant α-phase. Furthermore, the presence of a significant amount of EA-phase (mainly β-phase) in all electrospun PVDF samples is indicated by a strong exclusive absorption peak for β-phase at 1276 cm$^{-1}$ and an undetectable, rather weak shoulder for γ-phase at 1234 cm$^{-1}$ that appeared for all samples. No other characteristics absorption peaks of γ-phase at 776, 812 cm$^{-1}$ were detected. Consequently, the other characteristics peaks for β-phase appeared at 471, 1431 cm$^{-1}$, while absorption peaks at 512, 841 cm$^{-1}$ are assigned to both β and γ (β + γ) phase. This is because, both β and γ phase has similar chain conformation resulting in similar absorption band peaks [61, 62]. For instance, γ-phase shows absorption band at 512 cm$^{-1}$ which is very close to β-phase absorption peak at 510 cm$^{-1}$ [52, 63], whereas, it is believed that the strong absorption band at 841 cm$^{-1}$ is due to β-phase and the γ-phase appeared as a shoulder at 833 cm$^{-1}$ as common to both polymorphs [61, 64]. To further clarify the identification of 841/512 cm$^{-1}$ band, if both the exclusive absorption peak for β-phase at 1276 cm$^{-1}$ and γ-phase at 1234 cm$^{-1}$ are present, the 841/512 cm$^{-1}$ band is assigned as (β + γ) phase. The former phase is assigned to 841/512 cm$^{-1}$ band if one of them is present and the other is absent [49]. Thus, as 841 cm$^{-1}$ band represents both the electroactive β and γ phases in PVDF, the fraction of electroactive ($F_{EA}$) phase presents in a sample containing all α, β, and γ can be quantified from Lambert− Beer law given as [19, 65],

$$F_{EA} = \frac{A_{EA}}{\left(\frac{K_{EA}}{K_\alpha}\right)A_\alpha + A_{EA}} \times 100\% \qquad \text{Equation (3)}$$

where, $A_\alpha$ and $A_{EA}$ refer to the absorbance of α and electroactive phases (at 763 and 841 cm$^{-1}$), respectively. $K_\alpha$ and $K_{EA}$ represent the absorption coefficients with values of 6.1 × 10$^4$ and 7.7 × 10$^4$ cm$^2$ mol$^{-1}$ respectively. Although eq. (3) was originally used to quantify only one of the EA-phase only (β or γ) [19, 66, 67], it was later utilized for quantification of both β and γ-phases [49, 60]. A peak deconvolution method has been employed for 841 cm$^{-1}$ band to further quantify the individual β and γ phases, considering that the strong absorption band at 841 cm$^{-1}$ is due to β-phase, while the γ-phase appeared as a shoulder at 833 cm$^{-1}$ as common to both polymorphs [61, 64]. The following eq. (4a, 4b) are used to quantify relative proportion of β and γ phases in respect to the total electroactive phase ($F_{EA}$) present in each PVDF sample [60].

$$F(\beta) = F_{EA} \times \frac{A_\beta}{(A_\beta + A_\gamma)} \times 100\% \qquad \text{Equation (4a)}$$

$$F(\gamma) = F_{EA} \times \frac{A_\gamma}{(A_\beta + A_\gamma)} \times 100\% \qquad \text{Equation (4b)}$$

where, $A_\beta$ and $A_\gamma$ are the area under the deconvoluted curve centered at 841 cm$^{-1}$ (Shown in Figure 4a, 4b, 4c) for β and γ phases respectively. The calculated β and γ phase fractions are summarized in Table 1.

Another alternative method Cho *et al.* have used to calculate the relative content of a particular phase by sum of the intensity ratio of the absorption peaks associated to the phase to those α, β, and γ-phases and is given as [58],

$$\frac{I_\beta \; or I_\gamma}{I_\alpha+I_\beta+I_\gamma} = \frac{(I_{471}+I_{512}+I_{841}+I_{1276}+I_{1431}) \; or \; (I_{1234})}{(I_{489}+I_{532}+I_{614}+I_{796}+I_{976})+(I_{471}+I_{512}+I_{841}+I_{1276}+I_{1431})+(I_{1234})} \quad \text{Equation (5)}$$

The calculated β and γ phase fractions using eq. (5) are also summarized in Table 1. It should be noticed from Table 1 that the β-phase fraction estimated from eq. (4) has increased from 58.3 for 15-2:1 to 64.1% for 15_1:1 as acetone content is increased (namely DMF:Acetone ratio decreased), while the γ phase fraction changed from 3.5 to 4.8% respectively. This result is in good agreement with the calculated values from eq. (5), however it differs from the result that obtained from eq. (1) calculated using XRD data. When DMF:Acetone ratio was changed from 2:1 to 1:1, the β -phase fraction has increased from 45.4 to 63.6% in 20% PVDF, while it is not changed significantly in 25%. In contrast, the γ-phase fraction obtained from eq. (4) has significantly large values (18.5-19.9%) in 20%, yielding a contradictory result as compared to the values obtained from eq. (5). This could lead to the conclusion that because eq. (4) deals with deconvoluted curve for β -phase centered at the band 841 cm$^{-1}$ with a small shoulder of γ -phase at 833 cm$^{-1}$, it may introduce inaccuracies because γ-phase peak at 833 cm$^{-1}$ is not possible to fit well, as seen in Figure 4. Despite the fact that, some authors have employed this strategy, no fitting parameters were reported anywhere [49, 60].

The overall total EA-phase calculated from eq. (4) and (5) is presented in Figure 5. From Figure 5, it is clear that for any particular DMF:Acetone ratio as the PVDF concentration is increased, F$_{EA}$ (%) is increased. However, a maximum $F_{EA}$ = 82.1% is obtained for 20% when the acetone DMF:Acetone ratio is 1:1 that calculated from eq. (4). There is no significant deviation in the results obtained from eq. (4) and (5), suggesting more reliable method for calculating EA-phase. This also confirms that, PVDF concentration 20% is the critical concentration where DMF:Acetone plays important role in the nucleation and stabilization of electroactive β and γ phases.

**Morphology study**

A significant influence on the morphology (shape, diameter) of PVDF fiber due to variation in the concentration of PVDF and DMF:Acetone ratio can be found in the SEM micrographs of the samples. It is evident from the micrographs that low concentration of PVDF, i.e., 15% results in beaded fiber, however when the DMF:Acetone ratio is reduced from 2:1 to 1:1, the bead formation is also reduced. This is because when the concentration of PVDF is low the entanglement of the polymer macromolecular chain in the solution is not very strong [28, 68]. Furthermore, when a sufficiently large field is applied, the electrospinning process changes to electrosprayying [69]. However, the SEM micrographs show that a 15% PVDF concentration is viscous enough for electrospinning, whereas different volume ratios of binary solvent have varying entanglements and evaporation rates, affecting the extensional viscosity of the jet and resulting in diverse morphologies [70]. Moreover, when acetone concentration is increased (namely by reducing DMF:Acetone ratio), consequently the surface tension of the solution is also reduced creating a stable Taylor cone and a continuous jet starts forming due to effective stretching of the solution. This leads to reduction in bead formation as evident from sample 15_1:1. It is therefore confirmed that the bead formation is mainly related to solvent evaporation rate controlled by the concentration of acetone in the solvent while viscosity of the samples (15_1:1, 15_1.5:1, and 15_2:1) does not change much as further confirmed by their fiber diameter [53, 70, 71]. Table 2 summarized morphology study based on fiber diameter and nature of beads in the fibers. In all three samples of 15%, ultrafine fibers are produced having diameter ranging from 76 to 86 nm. Further, in PVDF concentration 20%, as the PVDF concentration is increased the viscosity of the solution is increased results in larger diameter (129-134 nm) fiber with almost negligible beads. Although, a very small number of beads are still found in 20_2:1, because of low evaporation rate of the solution, there are completely bead free fibers are obtained in 20_1:1. So, it can be concluded that, if a polymer solution jet is allowed to evaporate completely before collected, a bead-free fibers are formed [24]. In PVDF concentration 25%, large diameter (230-259 nm) with almost bead-free fibers can be seen. However, the viscosity is high enough to overshadow the influence of acetone concentration unless DMF:Acetone is 1:1 in 25_1:1. Therefore, it is speculated that, in 25_2:1 and even in 25_1.5:1, the solvent cannot be evaporated completely before collected in collector and small traces of liquid solvent still present. This results in wrinkle in the fiber due to further evaporation of the solvent. This was confirmed from the high magnification SEM micrograph shown in Figure 6. As can be seen from Figure 6, an almost uniform fibers with smooth surface are observed in all other samples of different PVDF concentrations 15 and 20% including 25_1:1 except 25:1.5:1, 25_2:1. Thus, it is confirmed that, acetone lower the surface tension while concentration of PVDF increase the viscosity which favors the formation of uniform and bead-free fibers which is mainly governed by the balance between viscosity and surface tension. Therefore, a suitable PVDF concentration and DMF:Acetone ratio is very much important for not only a desired morphology of the fibers but also to obtain the maximum EA-phase. The concentration lower than 15%, results in much beaded fibers even at any DMF:Acetone ratio, while PVDF concentration 15% with DMF:Acetone of 1:1 shows minimum numbers of beads. The DMF:Acetone ratio less than 1:1 was not used as it can dry up quickly results in clogging of the needle. Therefore, the most suitable combination was found for PVDF concentration 20% with DMF:Acetone ratio 1:1. It is believed that PVDF concentration 20% is just above the critical concentration and a DMF:Acetone ratio

1:1 is much favorable for complete solvent evaporation during electrospinning to obtain bead-free fibers [72, 73]. Although, a higher PVDF concentration 25% can produce bead-free fibers but solvent evaporation during electrospinning was not possible which could have a detrimental effect in EA-phase formation. It is expected that the macromolecular chain of the polymer solidifies under the high elongation rate during electrospinning to enhance the polymer crystallinity and consequently the EA-phase. If, however, fibers contain traces of solvent after collection, effective stretching in the air will not occur and therefore, orientation of the macromolecule are allowed to relax back resulting in α-phase [53]. As a consequence, the EA-phase in 25% is reduced as compared to 20%. So, intensive stretching of fibers leads to conversion of PVDF into β-phase, while a suitable solvent evaporation rate contributes to retain β-phase in solidified fibers.

**Conclusions**

The influence of electrospinning parameters such as PVDF concentration and DMF:Acetone ratio on the phase formation and morphology of PVDF nanofibers have been investigated. It is found that, viscosity and solvent evaporation rate of the solution played the most important role in obtaining the desired EA-phase and morphology. The content of acetone lowers the surface tension while concentration of PVDF increases the viscosity which favors the formation of uniform and bead-free fibers. Although evaporation rate is primarily responsible for bead formation at PVDF concentration ~ 15% (w/v), it is a delicate balance between viscosity and surface tension in the case of high concentrations of PVDF (25% w/v). Further, accurate identification of different PVDF phases (α, β, and γ) was possible by deconvoluting XRD and FTIR peaks. The study helps in finding a suitable PVDF concentration (20 % w/v) and DMF:Acetone ratio (1:1) for obtaining maximum β-phase and bead-free nanofibers and would be useful for understanding the actual mechanism for phase formation and morphology development. This would provide a guideline for further optimization of PVDF based nanocomposite nanofiber preparation in future.

## Acknowledgement

**Conflict of interest:** All authors declare there is no conflict of interest

# Figure caption

**Figure 1:** XRD patterns of PVDF samples with varying concentrations (15, 20, 25 %w/v) and DMF:Acetone ratios (1:1, 1.5:1, 2:1).

**Figure 2:** The deconvoluted peaks all the PVDF samples for the 2θ range 10-30º

**Figure 3:** FTIR spectra of PVDF samples with varying concentrations (15, 20, 25 %w/v) and DMF:Acetone ratios (1:1, 1.5:1, 2:1) shown for the frequency range 450-1500 cm$^{-1}$.

**Figure 4:** Deconvoluted curve for the band 841 cm$^{-1}$: (red) for γ -phase centered at 833 cm$^{-1}$; (green) for β -phase centered at 841 cm$^{-1}$; (blue) cumulative fit for PVDF concentration (a) 15, (b) 20, and (c) 25% with DMF:Acetone ratios: 1:1, 1.5:1 and 2:1.

**Figure 5:** Fraction of total electroactive phase F$_{EA}$ (%) at different DMF:Acetone ratio of 1:1, 1.5:1 and 2:1 for PVDF concentrations 15, 20, 25%.

**Figure 6:** SEM micrographs of PVDF samples with varying concentrations (15, 20, 25 %w/v) and DMF:Acetone ratios (1:1, 1.5:1, 2:1) is shown.

**Figure 7:** High magnification SEM micrographs of PVDF samples with varying concentrations (15, 20, 25 %w/v) and DMF:Acetone ratios (1:1, 1.5:1, 2:1) is shown.

# Table caption

**Table 1:** Summary of calculated crystallinity and β and γ phase fractions for different DMF:Acetone ratio of 1:1, 1.5:1 and 2:1 for PVDF concentrations 15, 20, 25%.

**Table 2:** Summary of morphological study of the nanofibers for different DMF:Acetone ratio of 1:1, 1.5:1 and 2:1 for PVDF concentrations 15, 20, 25%.

**FIGURE 1**

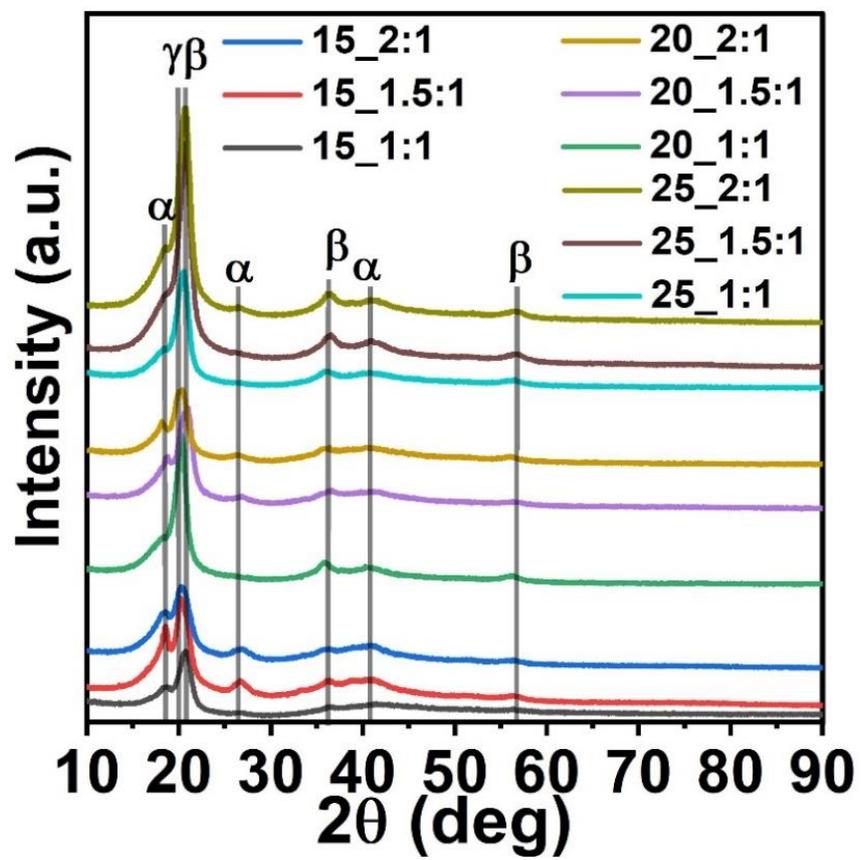

**FIGURE 2**

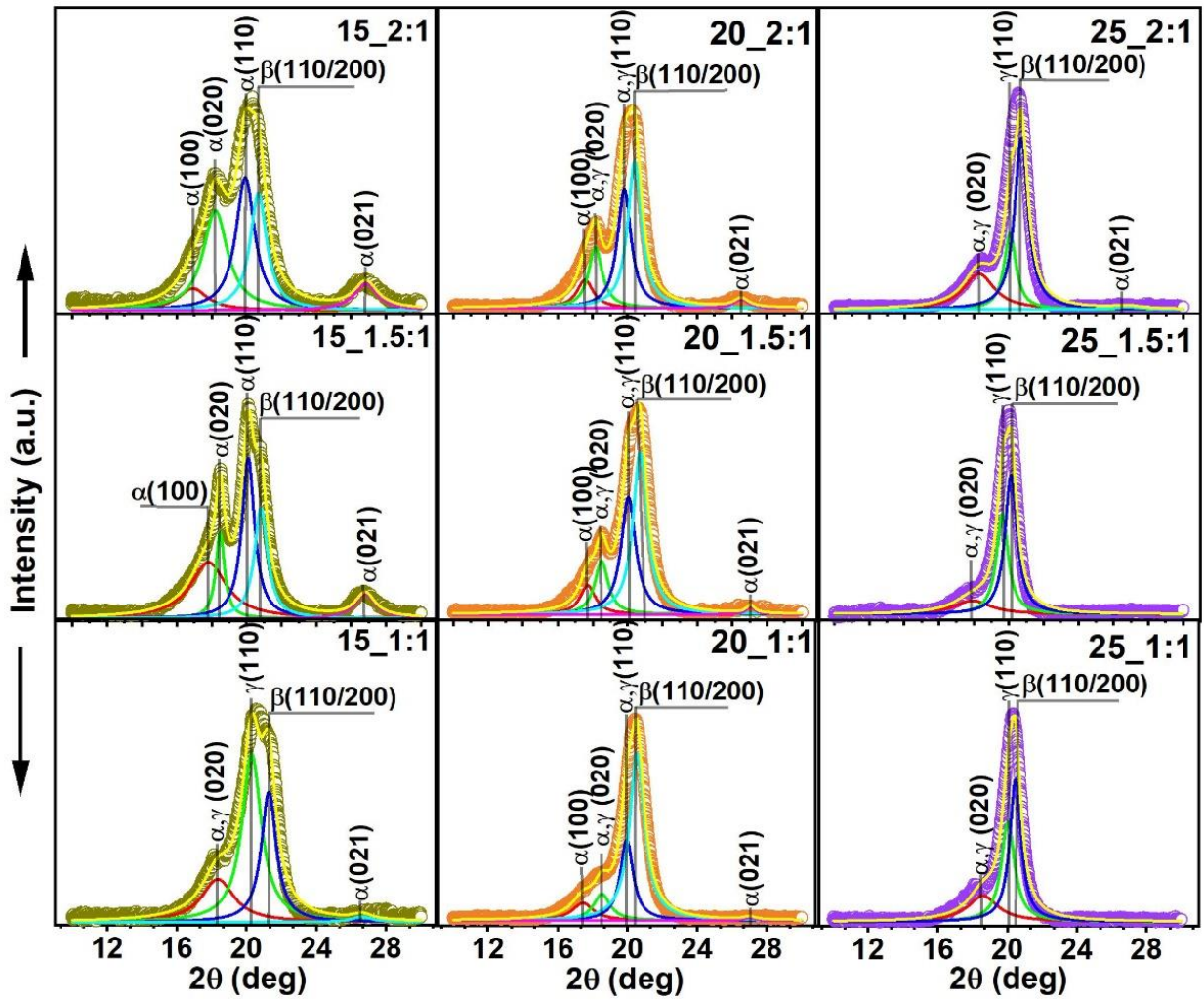

**FIGURE 3**

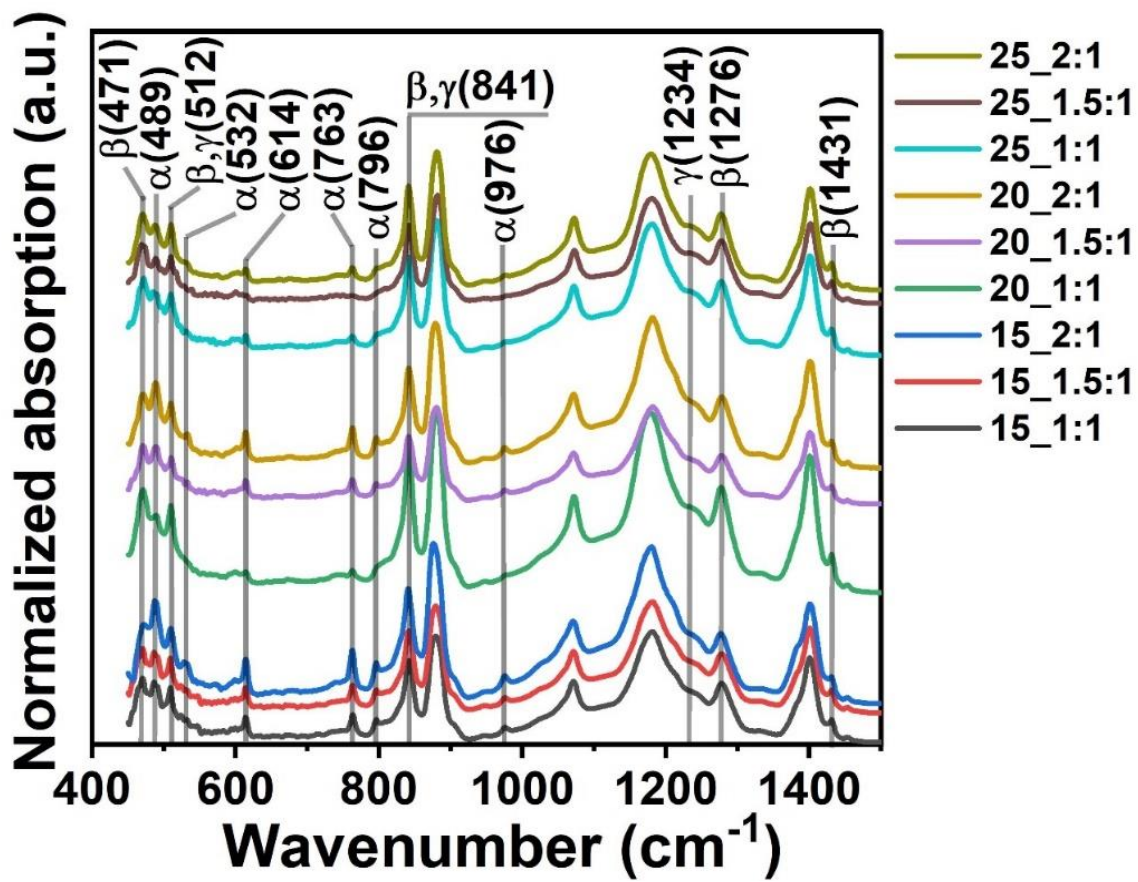

**FIGURE 4**

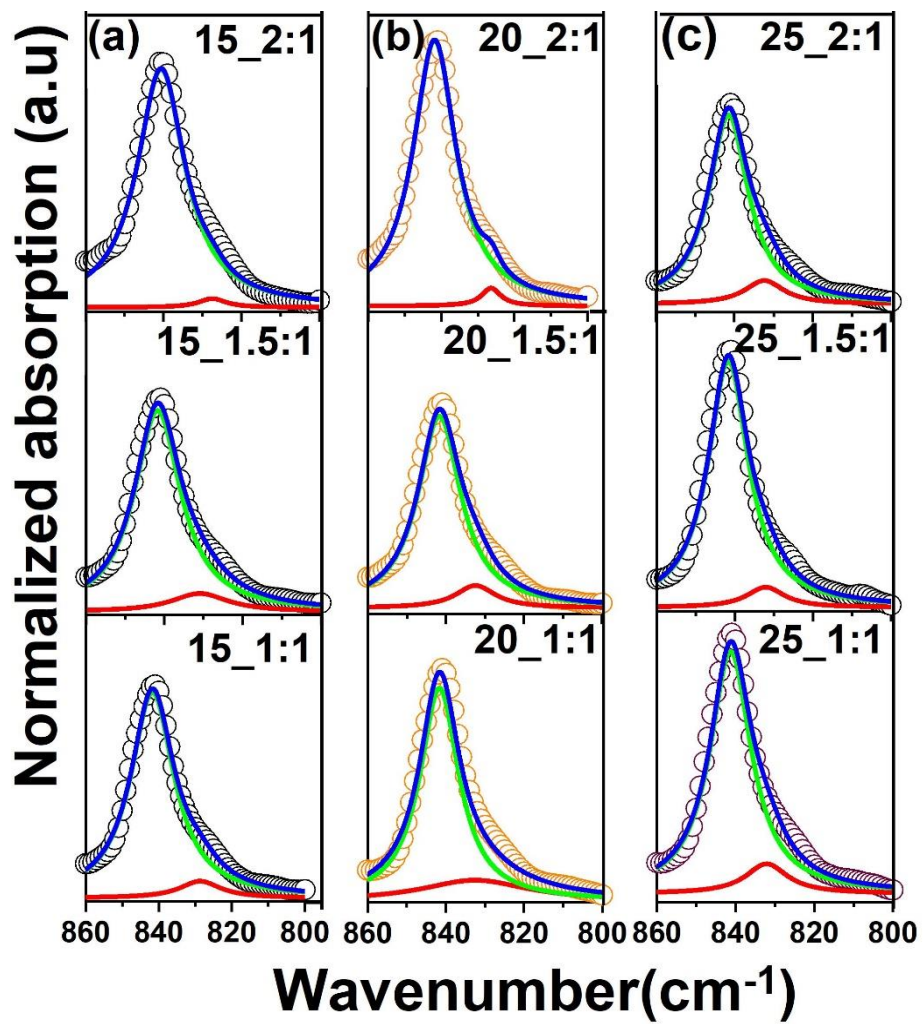

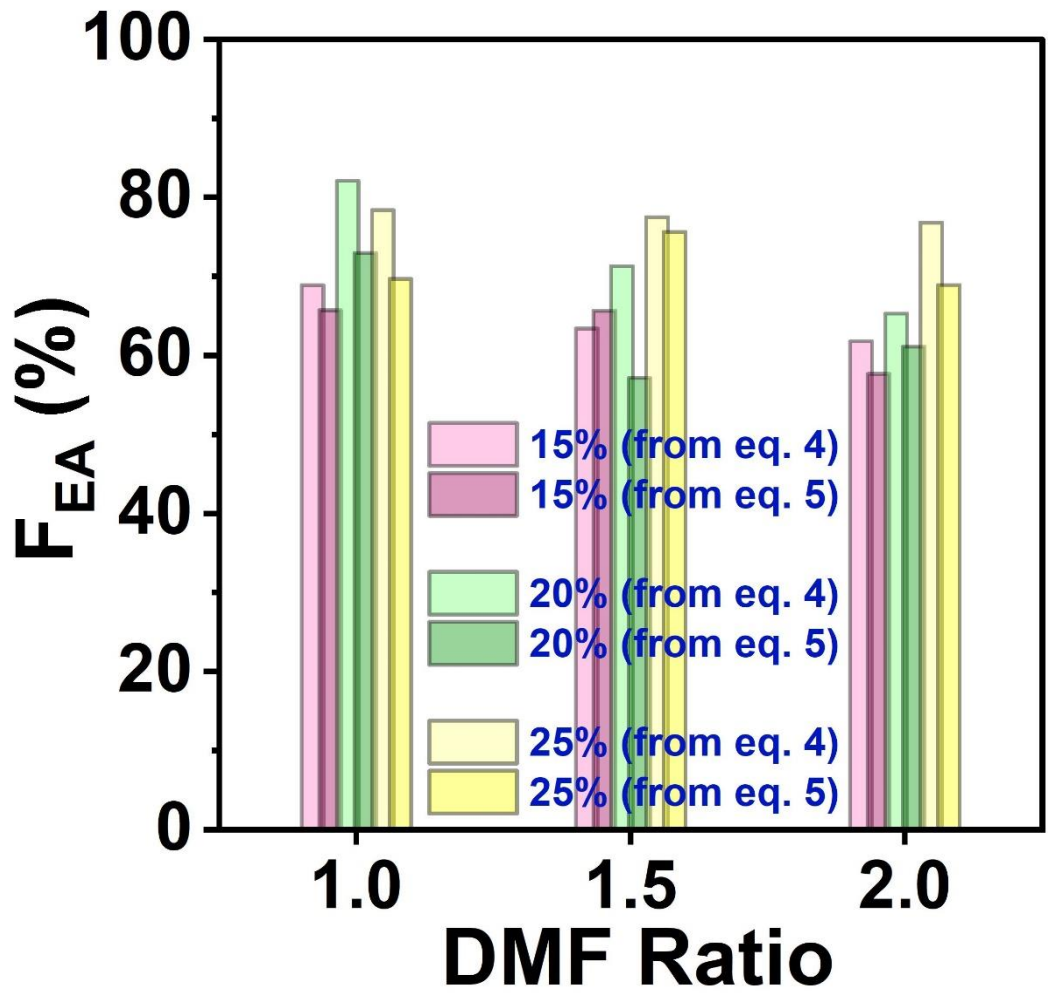

**FIGURE 6**

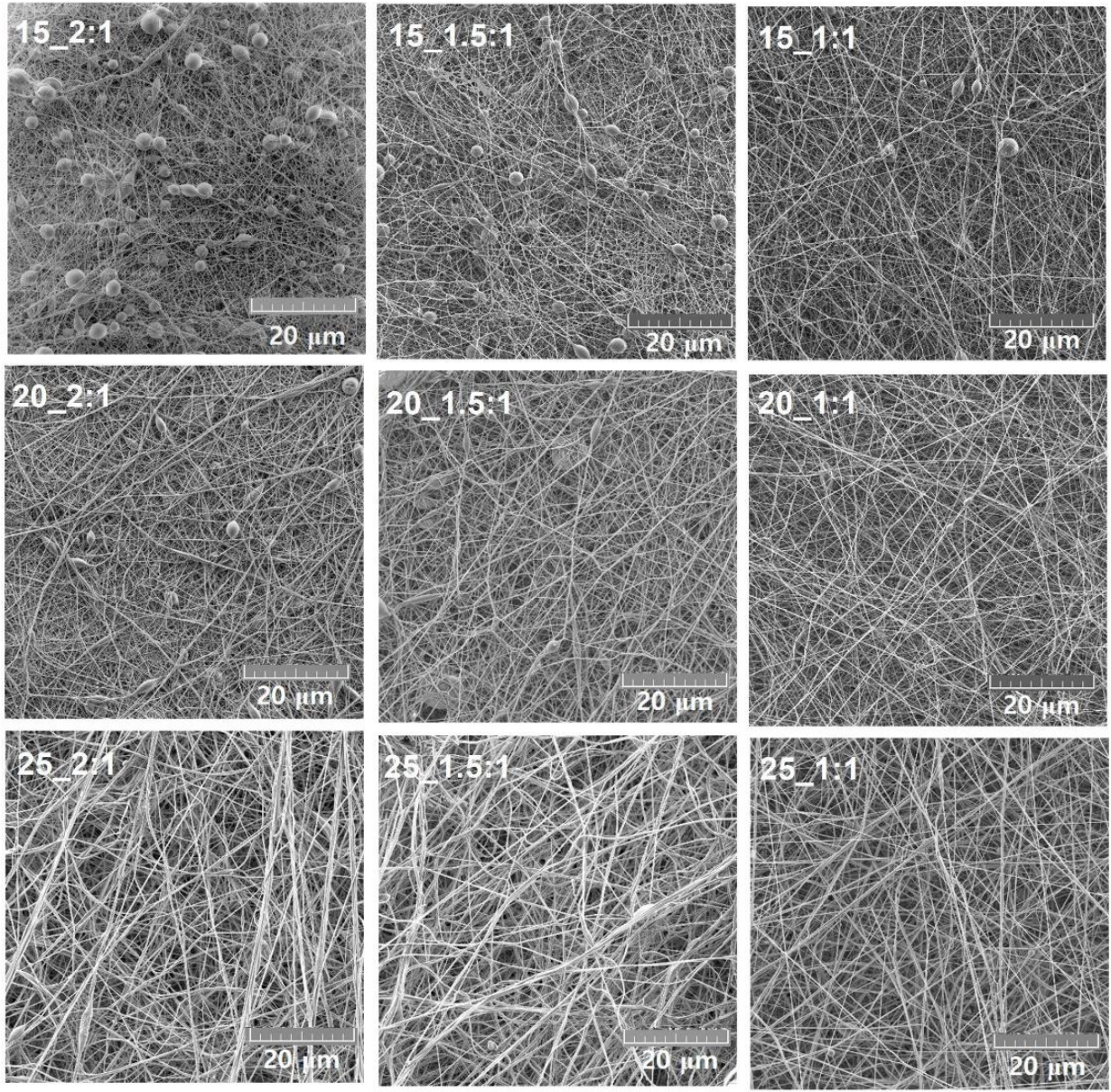

**FIGURE 7**

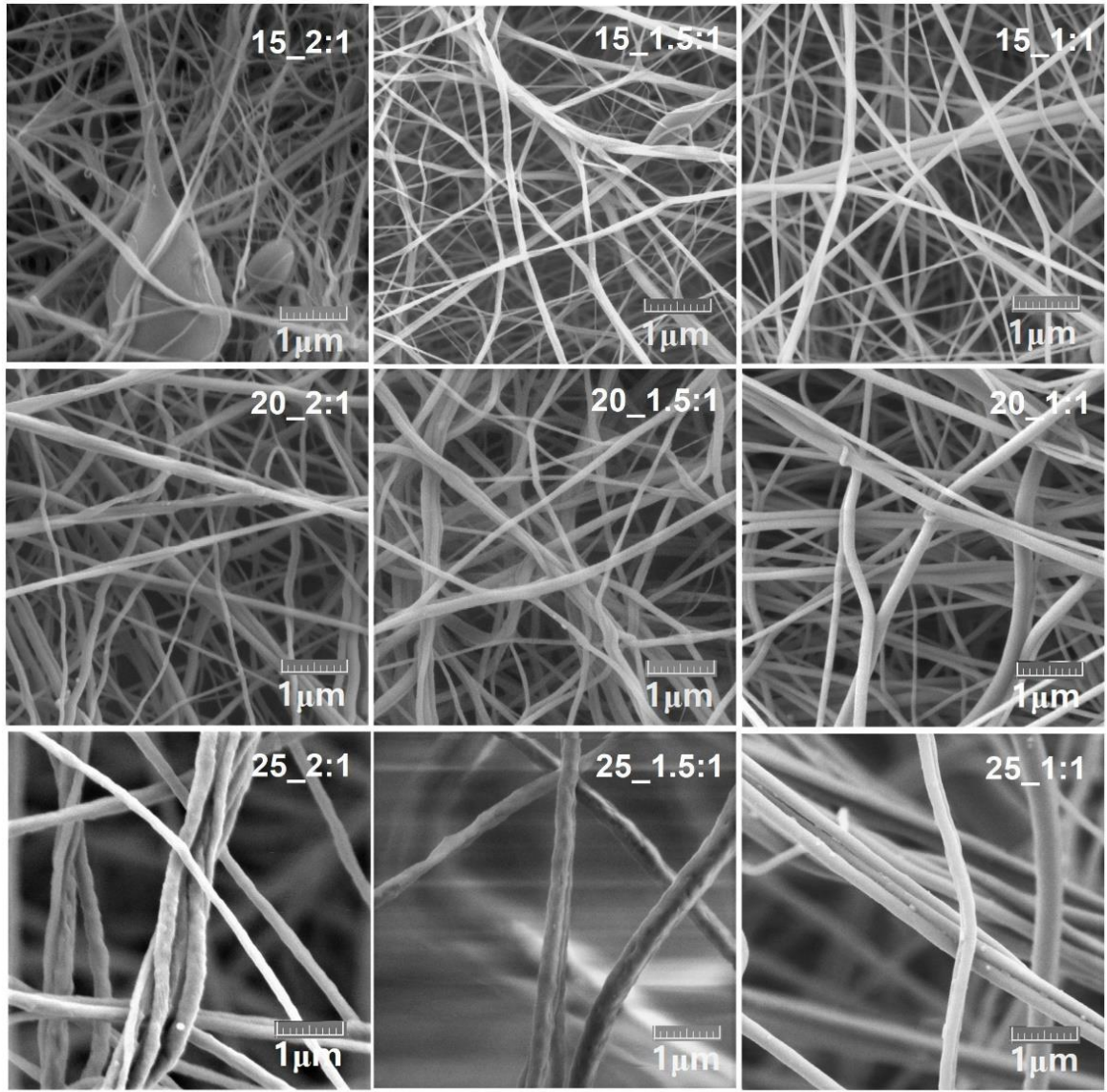

**Table 1:** Summary of calculated crystallinity and β and γ phase fractions for different DMF:Acetone ratio of 1:1, 1.5:1 and 2:1 for PVDF concentrations 15, 20, 25%.

| Sample id | Crystallinity% ($\chi_c$) From Eq. (2) | % of β-phase | | | % of γ-phases | | % EA-phases | |
|---|---|---|---|---|---|---|---|---|
| | | From Eq. (1) | From Eq. (4a) | From Eq. (5) | From Eq. (4b) | From Eq. (5) | From Eq. (4) | From Eq. (5) |
| 15_1:1 | 28.9 | 38.2 | 64.1 | 55.6 | 4.8 | 10.1 | 68.9 | 65.7 |
| 15_1.5:1 | 21.6 | 25.9 | 60.2 | 55.3 | 3.2 | 10.2 | 63.4 | 65.5 |
| 15_2:1 | 23.8 | 30.0 | 58.3 | 48.4 | 3.5 | 9.1 | 61.8 | 57.5 |
| 20_1:1 | 56.8 | 59.2 | 63.6 | 62.0 | 18.5 | 11.0 | 82.1 | 73 |
| 20_1.5:1 | 43.9 | 44.1 | 52.2 | 45.1 | 19.1 | 12.1 | 71.3 | 57.2 |
| 20_2:1 | 39.6 | 40.9 | 45.4 | 50.9 | 19.9 | 10.1 | 65.3 | 61.0 |
| 25_1:1 | 43.2 | 51.6 | 67.9 | 58.5 | 10.5 | 11.2 | 78.4 | 69.7 |
| 25_1.5:1 | 49.6 | 55.2 | 70.2 | 60.6 | 7.3 | 11.0 | 77.5 | 71.6 |
| 25_2:1 | 52.3 | 53.7 | 69.2 | 58.0 | 7.6 | 10.9 | 76.8 | 68.9 |

**Table 2:** Summary of morphological study of the nanofibers for different DMF:Acetone ratio of 1:1, 1.5:1 and 2:1 for PVDF concentrations 15, 20, 25%.

| Sample id | Diameter of fiber (nm) | Nature of beads |
|---|---|---|
| 15_1:1 | 80 ± 34 | Very small number, mostly spindle shape, smooth surface |
| 15_1.5:1 | 86 ± 43 | Small number, spherical and spindle shape, smooth surface |
| 15_2:1 | 76 ± 25 | Large number, spherical shape, smooth surface |
| 20_1:1 | 134 ± 57 | No beads, smooth surface |
| 20_1.5:1 | 129 ± 40 | Extremely small number, spindle shape, smooth surface |
| 20_2:1 | 130 ± 40 | Very small number, mostly spindle shape, smooth surface |
| 25_1:1 | 230 ± 73 | No beads, smooth surface |
| 25_1.5:1 | 259 ± 67 | No beads, wrinkle in surface |
| 25_2:1 | 241 ± 44 | No beads, wrinkle in surface |